\documentclass[aps,prd,twocolumn,groupedaddress]{revtex4}
\usepackage{graphicx}
\usepackage{dcolumn}
\usepackage{bm}
\begin{document}

\title{Shear viscosity of the quark matter}

\author{Masaharu Iwasaki, Hiromasa Ohnishi$^1$, and Takahiko Fukutome}
\email{miwasaki@cc.kochi-u.ac.jp}
\affiliation{Department of Physics, Kochi University, Kochi 780-8520, Japan \\
$^{1}$Institute of Materials Structure Science, KEK, Tsukuba 305-0801, Japan}
\email{ohni@post.kek.jp}

\date{\today}

\begin{abstract}
We discuss shear viscosity of the quark matter by using Kubo formula. The shear viscosity is calculated in the framework of the quasi-particle RPA for the Nambu-Jona-Lasinio model. We obtain a formula that the shear viscosity is expressed by the quadratic form of the quark spectral function in the chiral symmetric phase. The magnitude of the shear viscosity is discussed assuming the Breit-Wigner type for the spectral function.
\end{abstract}


\maketitle

\section{INTRODUCTION}

The existence of the quark gluon plasma (QGP), which is predicted by Quantum chromodynamics, has not been discovered in Nature. In order to produce such a new state of matter, the experimental investigation started at the Relativistic Heavy Ion Collider (RHIC). The data at RHIC, however, seems to reveal some unexpected properties of the high density matter produced in the experiment \cite{IA05}-\cite{KA05}. Namely it could be explained by a fluid model with small viscosity; it is almost perfect fluid. This fact encourages many researchers to calculate the transport coefficients of the quark matter \cite{HK85D}-\cite{MS04}. However their calculations are so complex that the numerical results are not settled completely. On the other hand, many studies by the lattice QCD have been in progress \cite{KW87}-\cite{NS05} Recently an approach from a very quite different field, black hole physics, has generated a great deal of interest. This theory is based on the AdS/CFT correspondence and predicts the lower bound of the shear viscosity of the quark matter \cite{KSS05}-\cite{HD05}.

It is the purpose of this paper to calculate the shear viscosity of the quark matter using the Nambu-Jona-Lasinio (NJL) model. It is known that free quark gas has infinite shear viscosity and the strong interaction between quarks lowers the value. We are interested in the physical origin of the small viscosity of a strongly interacting system. Hence we take up only the quark sector in the QGP in this paper. As for the calculation method of the viscosity, we use the correlation function method which is so-called Kubo formula \cite{K57}-\cite{Z74}.

In the next section, the Kubo formula for the shear viscosity is reviewed. Then it is calculated with the use of the quasi-particle RPA in \S 3. Finally assuming the Breit-Wigner type for the spectral function, the shear viscosity is evaluated and several discussions are given.

\section{Kubo Formula}

According to the Kubo formula, the shear viscosity $\eta(\omega)$ at temperature $T$ is given by
\begin{equation}
\eta(\omega) =\frac{1}{T}\int_{0}^{\infty}\mathrm{d}t\mathrm{e}^{i\omega
 t}\int\mathrm{d}{\bf r}(J_{xy}(\mathbf{r},t),J_{xy}(0,0)),
\end{equation}
where $J_{xy}$ is the $x,y$ component of the energy-momentum tensor of the quark matter. The correlation function in the right-hand side is defined by
\begin{equation}
(A,B) \equiv
 \beta^{-1}\int_{0}^{\beta}\mathrm{d}\lambda\langle\mathrm{e}^{\lambda
 H}A\mathrm{e}^{-\lambda H}B\rangle,
\end{equation}%
where $A$ and $B$ are operators of any physical quantity and $H$ denotes our Hamiltonian. The bracket, $\langle A \rangle = {\rm Tr}(A\mathrm{e}^{-\beta H})/{\rm Tr}\mathrm{e}^{-\beta H}$, means the thermal average at temperature $T$ ($\beta\equiv 1/T$). Using partial integration in the right-hand side of Eq.(1), the viscosity can be transformed into 
\begin{equation}
\eta(\omega) =\frac{i}{\omega}[\Pi^\mathrm{R}(\omega)-\Pi^\mathrm{R}(0)].
\end{equation}%
Here $\Pi^R (\omega)$ is a retarded Green's function defined by
\begin{equation}
\Pi^R (\omega) =-i\int_{0}^{\infty}\mathrm{d}t\mathrm{e}^{i\omega
 t}\int\mathrm{d}{\bf r}\langle[J_{xy}(\mathbf{r},t),J_{xy}(0,0)]\rangle,
\end{equation}%
where [ , ] in the integrand denotes the commutation relation. Noting that $(\Pi^R (\omega))^{*}=\Pi^R (-\omega)$, the (static) viscosity is reduced to
\begin{equation}
\eta\equiv \eta(\omega=0)=\left.-\frac{d}{d\omega}{\rm Im}\Pi^R (\omega)\right|_{\omega=+0}.
\end{equation}%
In order to calculate the above $\Pi^R (\omega)$, it is convenient to transform into the imaginary time (Matsubara) formalism. We introduce the following correlation function,
\begin{equation}
\Pi(i\omega_n) =-\int_{0}^{\beta}\mathrm{d}\tau\mathrm{e}^{{-i\omega_{n}\tau}}\int\mathrm{d}{\bf r}\langle T_\tau(J_{xy}(\mathbf{r},\tau) J_{xy}(0,0))\rangle,
\end{equation}%
where the Matsubara frequency is represented by $\omega_{n}=2\pi nT$ and $T_{\tau}$ means the (imaginary) time ordering operator. As is well known, the retarded Green's function $\Pi^R (\omega)$ is obtained by the analytic continuation: $\Pi^R (\omega) =\left.\Pi(i\omega_{n})\right|_{i\omega_{n}=\omega+i\delta}$.

We take the Nambu-Jona-Lasinio (NJL) model for the quark matter in this paper \cite{HK94}. The Lagrangian density is given by
\begin{equation}
\mathcal{L}=\bar\psi(i\gamma \cdot \partial-m)\psi +g[(\bar\psi\psi)^2+(\bar\psi i\gamma_{5}{\bf \tau}\psi)^2].
\end{equation}
Here $\psi$ is the field operator for quarks and the quark mass is assumed to be zero ($m=0$). Then the canonical energy-momentum tensor is read as
\begin{equation}
J_{xy} =\frac{i}{2}[\bar\psi\gamma^2\partial^1\psi-\partial^1\bar\psi\gamma^2\psi].
\end{equation}
If this expression is substituted into Eq.(6), the correlation function $\Pi$ is calculated according to the Wick theorem (Feynman diagram method). Our interaction is represented by the Feynman diagram Fig.1 where $\Gamma$ is $1$ (unit matrix) or $i\gamma_{5}{\bf \tau}$. Note that the four Fermi operators are composed of two pairs which are connected by a broken line.

\begin{figure}
 \includegraphics[width=\linewidth]{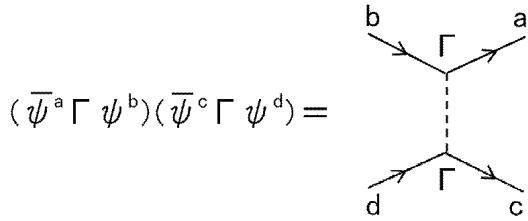}
 \caption{\label{fig:epsart}The diagram for the interaction with $\Gamma=1$ or $i\gamma_{5}{\bf \tau}$}
\end{figure}

\section{Quasi-particle RPA}

 Here we take the Fermi liquid theory and the random phase approximation (quasi-particle RPA). Namely the correlation function is approximated by the ring diagrams as shown in Fig.2. All the propagators must be the dressed ones because the free propagator gives rise to the infinity in the lowest term as shown in the last section. Therefore the ring approximation is broken down. In fact the quarks have strong correlation even in the chiral symmetric phase \cite{I02}. A similar calculation was done in a relativistic scalar field theory in the Ref.(30) where the shear viscosity was calculated by taking into account chain diagrams.

\begin{figure}
 \includegraphics[width=\linewidth]{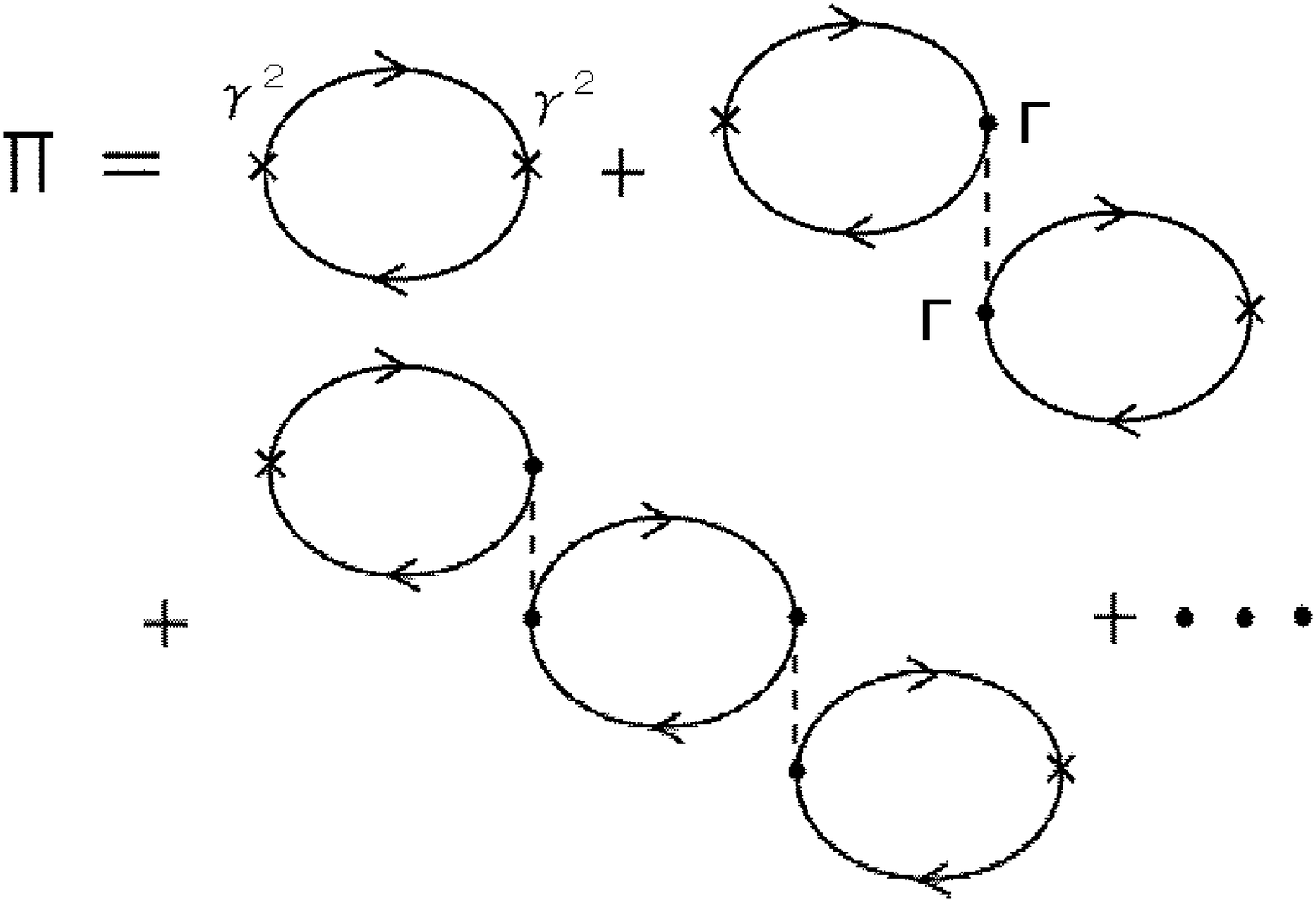}
 \caption{\label{fig:epsart}The ring diagrams for the correlation function $\Pi$.}
\end{figure}

The present approximation is also derived by the $1/N$ expansion used in QCD ($N=N_{c}$) \cite{TH74}-\cite{W79}. Let a diagram has $n$ vertices and $n_{l}$ quark loops. Let us consider the $N$-dependence of this diagram. Each quark loop gives rise to $N$ from the trace of color variable. The coupling constant of the NJL model is order of $1/N$ because the quark-gluon coupling constant is order of $1/\sqrt{N}$ in the $1/N$ expansion. Consequently this diagram is order of $N^{n_{l}-n}$. Therefore the leading diagrams are those with the maximum number of the loop for the fixed $n$. It is proved that $n_{l}\leq n+1$.

\noindent{\it proof}: Let us consider any diagram with $n$ vertices. If the number of the quark propagators is $n_{i}$, we have the relation, $2n_{i}=4+4n$. Each propagator belongs to only one loop necessarily according to the Wick theorem. Since each loop contains two or more propagators, the maximum number of the loop is realized that all the loops are composed of two propagators: $n_{l}=n_{i}/2=n+1$. The loop composed of one propagator (tadpole) is included in the dressed propagator. Thus it is evident that such diagrams correspond to the ring ones of Fig.2 and their magnitude is $O(N^{1})$.

For example, let us consider the first order diagrams with respect to the interaction shown in Fig.3. One has two loops (Hartree term) and the other has only one loop (Fock term) composed of four propagators. The former (a) is $O(N^{1})$ because it has one vertex and two quark loops. The latter (b) with one vertex and one loop is $O(1)$. Consequently the latter term should be neglected. In the same way the third diagram in Fig.2 is order of $N$ and so on.

\begin{figure}
 \includegraphics[width=\linewidth]{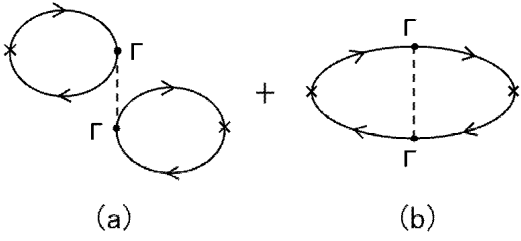}
 \caption{\label{fig:epsart}The first order diagrams: (a) Hartree term and (b) Fock term.}
\end{figure}

Now the correlation function is written as 
\begin{eqnarray}
\Pi(i\omega_n)
 &=&\frac{1}{\beta}\sum_l\int\frac{\mathrm{d}{\bf p}}{(2\pi)^3}{p_x}^2{\rm Tr}[{\gamma^2}G({\bf p},\omega_l+\omega_n){\gamma^2} \nonumber\\
 & & G({\bf p},\omega_l)]+\frac{2g}{\beta}\sum_{l,m}\int\frac{\mathrm{d}{\bf p}}{(2\pi)^3}\frac{\mathrm{d}{\bf q}}{(2\pi)^3} p_{x}q_{x} \nonumber\\
& & {\rm Tr}[{\gamma^2}G({\bf p},\omega_l+\omega_n){\Gamma}G({\bf p},\omega_l)]
\nonumber\\
& &{\rm Tr}[{\Gamma}G({\bf q},\omega_m-\omega_n){\gamma^2}G({\bf q},\omega_m)] \cdots ,
\end{eqnarray}%
where $G({\bf p},i\omega_n)$ is the dressed quark propagator. It should be noted that the dressed propagator is proportional to the linear combination of $\gamma^{\mu}$ in the chiral limit \cite{W00}. This leads to ${\rm Tr}[{\gamma^2}G{\Gamma}G]=0$ in the second term because the trace has an odd product of $\gamma$ matrices ($\Gamma=1$ or $i\gamma_{5}{\bf \tau}$). Thus the second term vanishes. Similarly the third term contains the same loop with odd product of $\gamma$ matrices so that it vanishes. The same calculation is realized in all the higher order terms in Fig.2. Consequently the correlation function $\Pi$ is reduced to only the first term in our present case.

In order to calculate the correlation function, let us follow the procedure taken for the calculation of the electrical conductivity in Ref[35]. We consider the spectral representation for the dressed propagator, which is written as\begin{equation}
G_{\alpha\beta}({\bf p},i\omega_l) =\int_{-\infty}^\infty
 \frac{\mathrm{d}\varepsilon}{2\pi} \frac{\rho_{\alpha\beta}({\bf p},\varepsilon)}{i\omega_l-\varepsilon}.
\end{equation}
We substitute this expression into the correlation function (Eq.(8)) and the summation over Matsubara frequency is replaced by the contour integral:
\begin{eqnarray}
S&\equiv &T\sum_l {\rm Tr}[{\gamma^2}G({\bf p},i\omega_l+i\omega_n){\gamma^2}G({\bf p},i\omega_l)] \nonumber \\
&=&-\int_C \frac{\mathrm{d}z}{2\pi i} n(z) {\rm Tr}[G({\bf p},z)\gamma^2 G({\bf p},z+i\omega_n)\gamma^2],
\end{eqnarray}
where $n(z)=(1+e^{\beta z})^{-1}$ is the Fermi distribution function. The contour $C$ is divided into three pieces as shown in Fig.4, because the integrand has branch cuts on the two lines $z=\varepsilon$ and $z=\varepsilon-i\omega_n$ where $\varepsilon$ is real (Note that poles of $n(z)$ do not lie on the two branch cuts).

\begin{figure}
 \includegraphics[width=\linewidth]{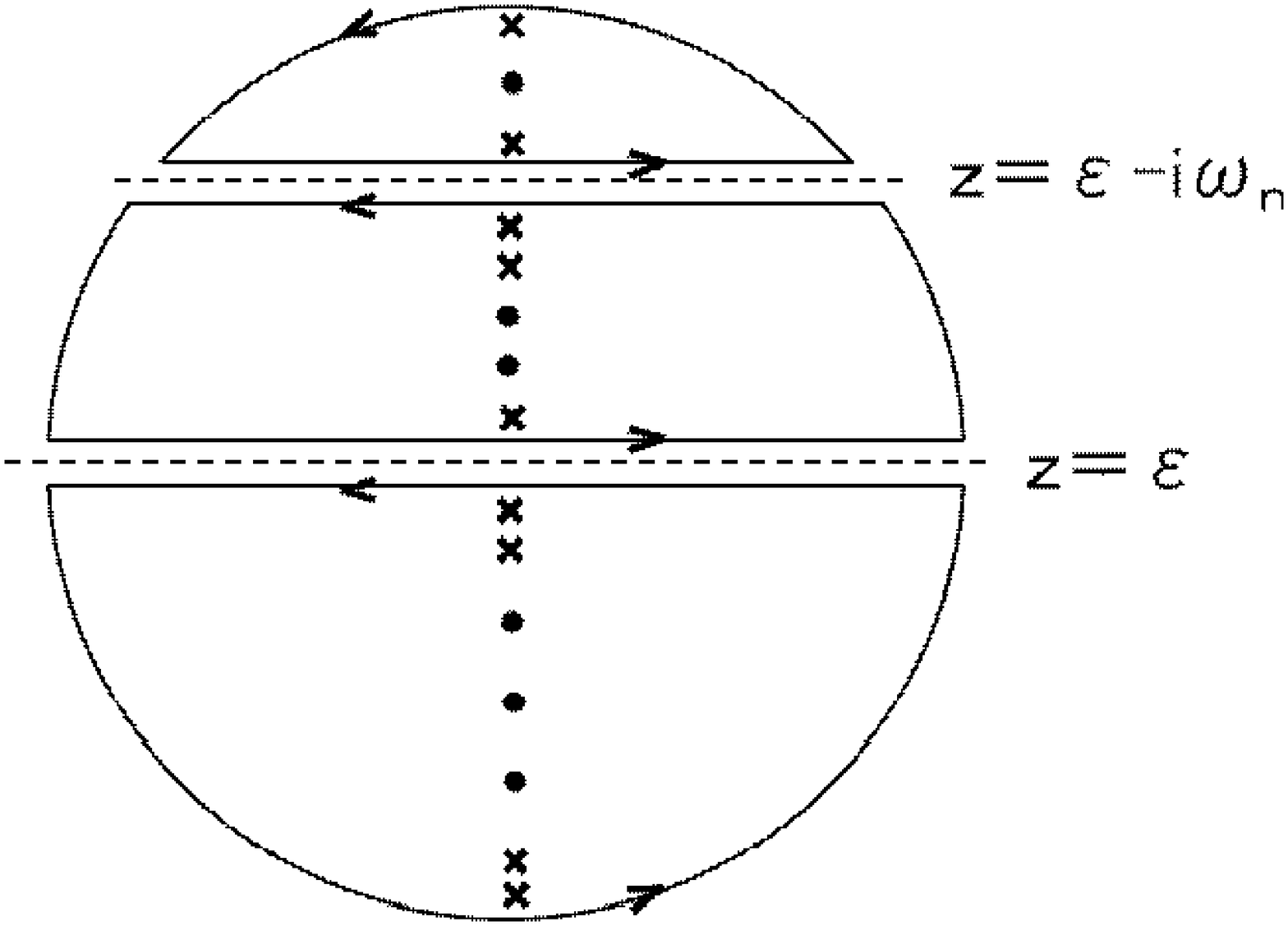}
 \caption{\label{fig:epsart}The contour for the calculating the integral of $z$ in Eq.(11).}
\end{figure}

Since the integral along the large circle vanishes, it is rewritten as
\begin{eqnarray}
S&=&-\int_{-\infty}^{\infty}\frac{\mathrm{d}\varepsilon}{2\pi
  i}n(\varepsilon){\rm Tr} [G(\varepsilon+i\delta)\gamma^2 G(\varepsilon+i\omega_n)\gamma^2 \nonumber\\
 & & -G(\varepsilon-i\delta)\gamma^2 G(\varepsilon+i\omega_n)\gamma^2 +G(\varepsilon-i\omega_n)\gamma^2 \nonumber\\
 & &G(\varepsilon+i\delta)\gamma^2-G(\varepsilon-i\omega_n)\gamma^2 G(\varepsilon-i\delta)\gamma^2], \nonumber
\end{eqnarray}%
where $\delta$ is an infinitesimal positive number introduced in order to avoid the branch cuts. Here the argument ${\bf p}$ is abbreviated for convenience. Noting the relation, $G(\varepsilon+i\delta)-G(\varepsilon-i\delta)=-i\rho(\varepsilon)$, the above equation is expressed as follows:
\begin{equation}
S = \int_{-\infty}^{\infty} \frac{\mathrm{d}\varepsilon}{2\pi} n(\varepsilon) {\rm Tr}[(G(\varepsilon+i\omega_n)+G(\varepsilon-i\omega_n))\gamma^2\rho(\varepsilon)
\gamma^2].
\end{equation}%
Substituting the spectral representation (10) again, we get
\begin{eqnarray}
S&=& \int_{-\infty}^{\infty} \frac{\mathrm{d}\varepsilon}{2\pi} \int_{-\infty}^{\infty} \frac{\mathrm{d}\varepsilon'}{2\pi} n(\varepsilon)\left(\frac{1}{\varepsilon+i\omega_{n}-\varepsilon'}+\frac{1}{\varepsilon-i\omega_{n}-\varepsilon'} \right) \nonumber\\
& &{\rm Tr}[\rho(\varepsilon')\gamma^2\rho(\varepsilon)\gamma^2].
\end{eqnarray}
By the analytic continuation $i\omega_{n}\longrightarrow \omega+i\delta$, the imaginary part of the above equation becomes
\begin{equation}
\mbox{Im}S = \int_{-\infty}^{\infty} \frac{\mathrm{d}\varepsilon}{2\pi}
 \frac{1}{2} (n(\varepsilon+\omega)-n(\varepsilon)) {\rm Tr}[\rho(\varepsilon+\omega)\gamma^2\rho(\varepsilon)\gamma^2],
\end{equation}%
because ${\rm Tr}[\rho(\varepsilon')\gamma^{2}\rho(\varepsilon)\gamma^{2}]$ is real. As a result we obtain an expression for the shear viscosity which is related to the quark spectral function,
\begin{equation}
\eta = -\frac{1}{2}\int_{-\infty}^{\infty}
 \frac{\mathrm{d}\varepsilon}{2\pi}
 \int\frac{\mathrm{d}{\bf p}}{(2\pi)^3}{p_x}^2\frac{\partial n}{\partial \varepsilon}{\rm Tr}[\rho({\bf p},\varepsilon)\gamma^2\rho({\bf p},\varepsilon)\gamma^2].
\end{equation}
This equation means that the calculation of the shear viscosity is reduced to that of the spectral function. The factor of ${\partial n}/{\partial \varepsilon}$ means that the main contribution comes from the neighborhood of the Fermi surface. Note that the information of the Fermi surface (chemical potential) is contained in the quark spectral function.

\section{Numerical Results}

Now we are in a position to discuss the quark spectral function when the chiral symmetry is restored. According to Weldon \cite{W00}, the general forms of the self-energy for the massless quark is expressed by
\begin{equation}
\Sigma(p)=\Sigma_{+}\gamma^{0}\Lambda_{+}+\Sigma_{-}\gamma^{0}\Lambda_{-},
\end{equation}
using the projection operator $\Lambda_{\pm }=(1\pm \hat{\bf p}{\bf r}$) and $p_{0}\equiv \varepsilon+\mu$ ($\mu$: the chemical potential). Then the retarded (advanced) Green's function are given by
\begin{eqnarray}
G^{R}(p) &=& \frac{\gamma^{0}\Lambda_{+}}{p^{0}_{+}-p-\Sigma_{-}}+\frac{\gamma^{0}\Lambda_{-}}{p^{0}_{+}+p-\Sigma_{+}}, \nonumber \\
G^{A}(p) &=& \frac{\gamma^{0}\Lambda_{+}}{p^{0}_{+}-p-\Sigma_{-}^{*}}+\frac{\gamma^{0}\Lambda_{-}}{p^{0}_{+}+p-\Sigma_{+}^{*}}.
\end{eqnarray}%
The spectral function is given by
\begin{equation}
\rho(p)=i[G^{R}-G^{A}]=\rho_{+}\gamma^{0}\Lambda_{+}+\rho_{-}\gamma^{0}\Lambda_{-},
\end{equation}
where the $\rho_{\pm}$ is defined by
\begin{equation}
\rho_{\pm}(p)\equiv \frac{-2\Im \Sigma_{\mp}}{(p^{0}_{+}\mp p-\Re \Sigma_{\mp})^2+(\Im \Sigma_{\mp})^2}.
\end{equation}
Thus we get a spectral function of the Breit-Wigner type. 

As for the calculation of the self-energy $\Sigma$, the coupling to the soft mode \cite{HK85} or the plasmino \cite{BP92}-\cite{B96} might be probable. Before that it is instructive to calculate the shear viscosity by using a simple form,
\begin{equation}
\Sigma_{\pm}(p)=M-i\Gamma,
\end{equation}
where $M$ and $\Gamma$ are positive constants and regarded as free parameters. Since the quasi-particles near the Fermi surface contribute main part of the integral in Eq.(15), this replacement is not so bad. The real part $M$ is taken to be zero since it essentially shifts the chemical potential. The imaginary part $\Gamma$ means the inverse of the mean free path (or the inverse of the life time) of the quasi-particle. The numerical calculation was done with the temperature $T=200{\rm MeV}$ and the chemical potential $\mu=10{\rm MeV}$ for convenience. The shear viscosity is shown in Fig.5 as a function of $\Gamma$ and the ratio of the shear viscosity to the entropy density is also shown in Fig.6.

\begin{figure}
 \includegraphics[width=\linewidth]{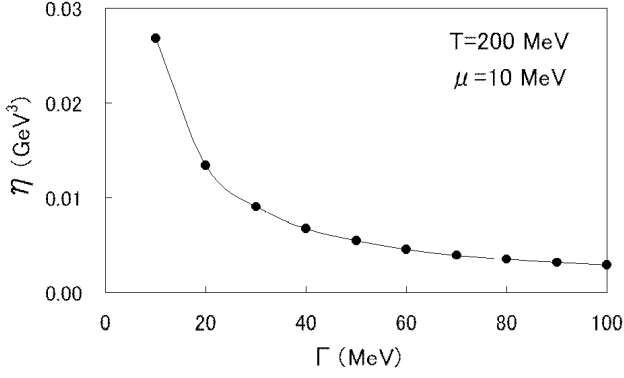}
 \caption{\label{fig:epsart}The shear viscosity as a function of the imaginary part of the self-energy at $T=200{\rm MeV}$ and $\mu=10{\rm MeV}$}
\end{figure}
\begin{figure}
 \includegraphics[width=\linewidth]{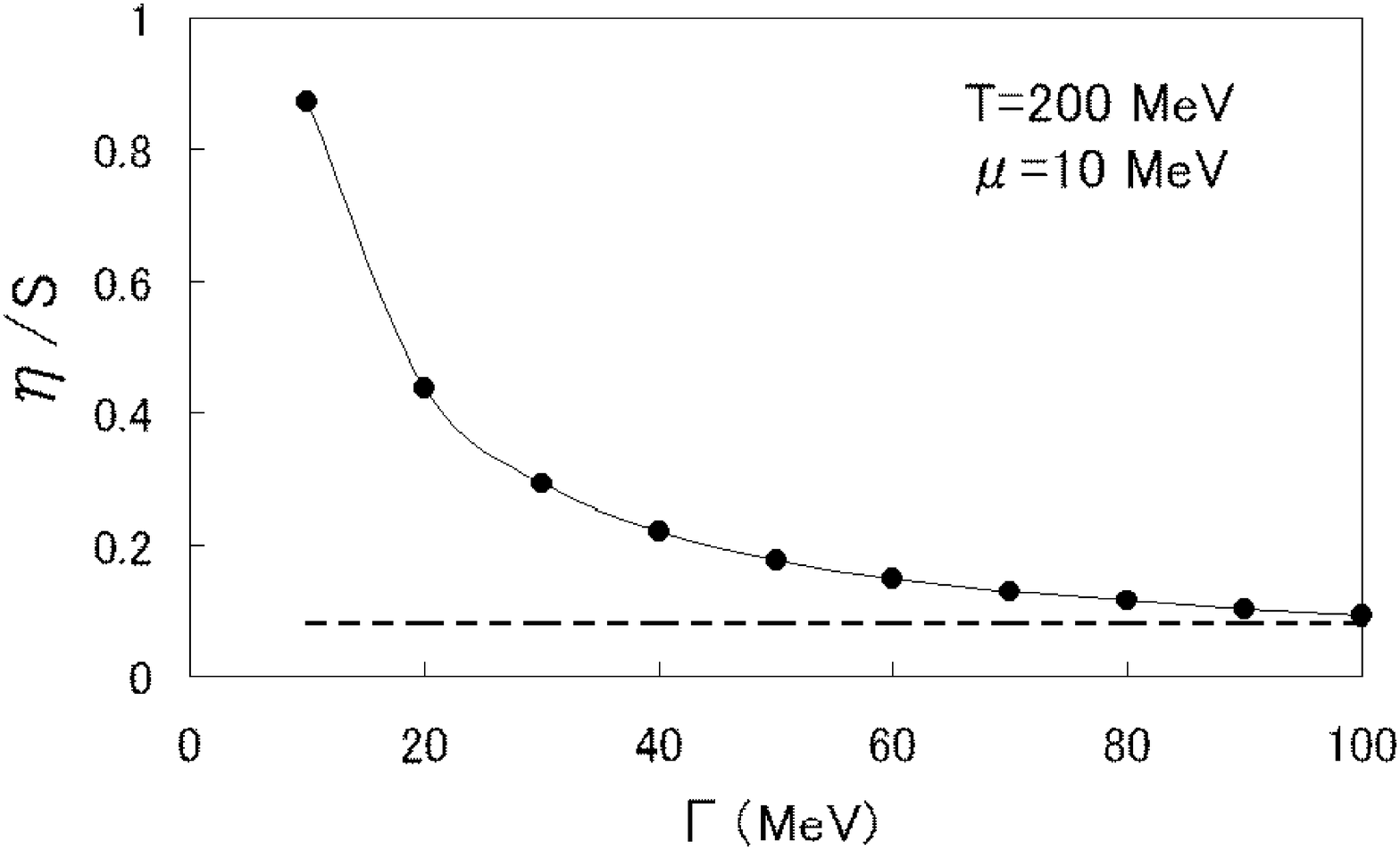}
 \caption{\label{fig:epsart}The ratio of the shear viscosity to the entropy as a function of the imaginary part of the self-energy at $T=200{\rm MeV}$ and $\mu=10{\rm MeV}$}
\end{figure}

 These figures evidently show that the viscosity is a rapidly increasing function of the mean free path, $\Gamma^{-1}$. The viscosity seems to diverge at $\Gamma=0$ where the quark matter becomes free gas (the mean free path is infinite). This is understood by noting that $(\Gamma/(x^2+\Gamma^2))^2\rightarrow (\pi\delta(x))^2$ as $\Gamma\rightarrow 0$. As the $\Gamma$ becomes larger, the viscosity decreases rapidly. This is also apparent because $(\Gamma/(x^2+\Gamma^2))^2\rightarrow 0$ as $\Gamma\rightarrow \infty$. This $\Gamma$ dependence of the viscosity is consistent with the classical expression of the viscosity of gas: $\eta=\rho vl/3$ ($\rho$=density, $v$=velocity and $l$= mean free path of the particle). The ratio $\eta/s$ of water in the normal state is about $30$. Therefore the quark matter is a perfect fluid under strongly interacting state ($\Gamma>1{\rm MeV}$). The lower bound predicted by the superstring theory is drawn by the dashed line in the Fig.6.

 Next let us comment the evaluation of the $\Gamma$ by considering the coupling between the quark and the soft mode as shown in Fig.7 \cite{HK94},\cite{85}. The dashed line denotes for the propagator of the soft mode. As an example, we show here $\Gamma(p=0,\omega)$ shown in Fig.8 at $T=200{\rm MeV}$ and $\mu=10{\rm MeV}$. (The critical temperature is $T_{c}=165{\rm MeV}$)
\begin{figure}
 \includegraphics[width=\linewidth]{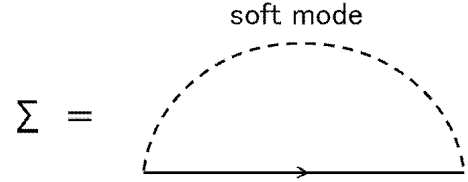}
 \caption{\label{fig:epsart}The self-energy diagram of the quark which is caused by the coupling to the soft modes.}
\end{figure}
\begin{figure}
 \includegraphics[width=\linewidth]{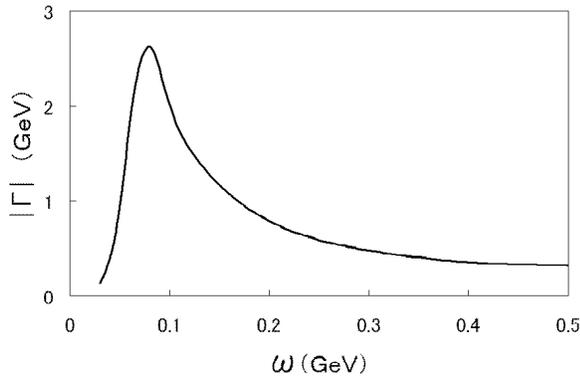}
 \caption{\label{fig:epsart}The calculated $\Gamma(p=0,\omega)$ at $T=170{\rm MeV}$ and $\mu=10{\rm MeV}$}
\end{figure}
This figure suggests that the quark matter has strong correlation even in the chiral symmetric phase: the $\Gamma$ is of order $0.1{\rm GeV}$. According to Fig.5, the shear viscosity would be small enough. Of course the detailed calculation of the shear viscosity itself is left in the future.
 
In conclusion, we have calculated the shear viscosity of the quark matter with the use of the Kubo formula. Assuming the quasi-particle RPA, we have obtained a simple formula for the shear viscosity, which is expressed by the quadratic form of the quark spectral function in the chiral symmetric phase. The magnitude of the shear viscosity is expected to be small in the strongly interacting system.

\begin{acknowledgments}
The authors would like to thank Professors S. Sakai (Yamagata University), M.Asakawa (Osaka University), Y.Tsue and K.Iida (Kochi University) for valuable comments and discussions. They also thank the Yukawa Institute for Theoretical Physics at Kyoto University. Discussions during the YITP workshop YITP-W-04-07 on Thermal Quantum Field Theories and Their Applications and the YKIS2006 on "New Frontiers QCD" were very useful to start and complete this work. 
\end{acknowledgments}

\end{document}